\documentclass[10pt,onecolumn,preprintnumbers,amsmath,amssymb]{revtex4}
\usepackage{graphicx,rotate}
\usepackage{float}
\usepackage{subfig}
\usepackage{color}
\usepackage{ulem}

\begin{document}

\title{Minimal length, maximal momentum and stochastic gravitational waves spectrum generated from cosmological QCD phase transition}
\author{Mohamed Moussa}
\email{mohamed.ibrahim@fsc.bu.edu.eg}
\affiliation{Physics Department, Faculty of Science, Benha University, Benha 13518, Egypt}
\author{Homa Shababi}
\email{h.shababi@scu.edu.cn}
\affiliation{Center for Theoretical Physics, College of Physical Science and Technology,
Sichuan University, Chengdu 610065, P. R. China}
\author{Anisur Rahaman}
\email[email: ]{manisurn@gmail.com}
\affiliation{Hooghly Mohsin College, Chinsura, Hooghly, India}
\author{Ujjal Kumar Dey}
\email{ujjal@iiserbpr.ac.in}
\affiliation{Department of Physical Sciences, Indian Institute of Science Education and Research, Berhampur 760010, India}

\begin{abstract}
We investigate thoroughly the temporal evolution of the universe temperature as a function of the Hubble parameter associated with the Stochastic Gravitational Wave (SGW), that formed at the cosmological QCD phase transition epoch to the current epoch, within the Generalized Uncertainty Principle (GUP) framework. Here we use GUP version which provide constraints on the minimum measurable length and the maximum
observable momentum, that characterized by a free parameter $\alpha$. We study the effect of this parameter on the SGW background. We show that the effect can slightly enhance the SGW frequency in the lower frequency regime which might be important in the detection of SGW in the future GW detection facilities.

\end{abstract}

\maketitle

\section{Introduction}
The recent detection of gravitational wave (GW) from the merger of black holes~\cite{LIGO} and its verification on repeated occasions has opened the way to a new era turning Einstein's prediction into a physical reality. It attracted high attention of physicists on both theoretical and observational sides. The recent development of technology and the enriched knowledge of fabrication of detectors with high sensitivity has enabled the LIGO scientific collaboration to capture high frequency ($10-10^3$ Hz) gravitational waves from the compact binary spiral~\cite{LIGO, VIR}. For the detection of low-frequency GW ($10^{-5}-1$ Hz) signal from the source like a supernova, the eLISA space probe has also been in a state of final preparation \cite{AKL}.
The stochastic background of gravitational waves is expected to arise from a superposition of a large number of unresolved gravitational-wave sources of astrophysical and cosmological origin which are considered as the potential sources of the QCD and electroweak cosmological phase transitions at the earliest epochs in the evolution of the universe. There has been great interest in detecting or constraining the strength of SGW that may have been produced by a variety of processes in the early universe, including inflation. The stochastic gravitational-wave background (SGWB) is by far the most difficult source of gravitational radiation to detect~\cite{AKL}. At the same time, it is the most interesting and intriguing since they carry profound information about the early stage of the universe. Therefore, mathematical modelling of the source in this situation is quite relevant and beneficial as well.
It has been shown that the QCD and electroweak cosmological phase transitions which lasted for adequate duration could be a potential source of very low-frequency SGWB~\cite{WIT, CJH, CJHO, MST}. In the context of the standard model of particle physics there are at least two types of phase transitions; the first one at $T \sim 100$ GeV, called electro-weak phase transition due to spontaneous symmetry breaking, and the second one at $T \sim 0.1$ GeV, known as the QCD phase transition due to the breaking of the chiral symmetry. These transitions although argued as a cross-over, the physics beyond the standard model entails that strong first-order transition at the QCD scales is possible~\cite{10,11,12,13,14,15,16,17,18,19,20}. The exact value of the critical temperature of this transition however is not settled unambiguously. The non-perturbative study of the hadronization process, relevant to strong QCD phase transition, is focused on reaching the relevant equation of state (EoS) governing the two different phases, quark-gluon plasma (QGP) and hadronic gas (HG). The recent results from QCD lattice studies, suggests that in the presence of strong interactions, the celebrated pressure-energy relation $P=\frac{1}{3}\rho$ does not satisfy in the radiation dominated epoch~\cite{DeTar}. It acquires a modification since the interaction measure is marked by the trace anomaly that is expected to lead to some exciting cosmic results such as the prediction of Weakly Interacting Massive Particles (WIMPs), and pure glue lattice QCD calculations~\cite{Hajkarim}.
Some interesting investigations are in progress in this direction \cite{SAN,SCAP}. Trace anomaly has been considered solemnly in \cite{SAN}, energy density is computed from trace anomaly to deduce the equation of state that emerged from parametrization of the pressure due to $u$, $d$, and $s$ quarks and the gluons. The remarkable differences that have been noticed there is that the rate of expansion of the universe is decreased, however, the gravitational wave signal is increased by almost 50\% with a redshift of the peak frequency to the current time by 25\%. In~\cite{SCAP}, a fractional energy density and redshift of the peak frequency at the current time of SGW is investigated using an effective QCD equation of state of three quark flavors $u$, $d$, and $s$ including chemical potential taking finite temperature effect into account where it is found that the frequency and amplitude of SGW signal present today, get enhanced with an increase in the chemical potential.
On the other hand, over the past decades, many attempts have been made to study various physical phenomena at the Planck scale. The natural cut-off momentum and minimal length, that produced in GUP theories, tackled some issues in physics such as appearance of a finite values of vacuum energy density in quantum field theory \cite{a}.
In cosmology, it is proved that GUP can fabricate an acceleration in the early universe but inhibits the undying acceleration at later time and turns it into deceleration \cite{b,c}. This result points out that GUP forfeits its effect with  later time evolution of the universe.
Then a cosmological implication of GUP are widely examined within early universe in order to solve a dilemma of the dark matter and dark energy, see for example \cite{d,e,f,g}.
One of the most important challenges of GUP theories is the absence of experimental evidence assure the presence of quantum gravity and its effect. For instance minimum length and cut-off energy, which are most important consequence of GUP theories, are not confirmed yet. Over the past decades, much effort is being made to test and confirm these hypotheses \cite{h,i,j}.
To capture the Planck scale effects, GUP is an important tool. So GUP modified equation of state will be reasonable to use in the study of the SGWB power spectrum. In this context, the GUP modified equations of state have been used to study the SGWP power spectrum in~\cite{SKHO, HOMA}, and the possibility of detection of the SGWB signal has been discussed. In this paper, we intend to investigate the power spectrum of SGWB with linear quadratic GUP to study qualitatively whether the detection is more probable in the current epoch.

%%%%%%%%%%%%%%%%%%%%%%%%%%%%%%%%%%%%%%%%%%%%%%%%%
\section{GUP modification in photons entropy}
\label{sec:gupMod}
%%%%%%%%%%%%%%%%%%%%%%%%%%%%%%%%%%%%%%%%%%%%%%%%%
At high-energy physics, close to the Planck scale, the effects of gravity become so important that it would lead to discreteness of the spacetime. In this vein, several approaches to quantum gravity such as string theory~\cite{Veneziano,Amati1,Amati2,Gross,Konishi} noncommutative geometry \cite{Capozziello3}, loop quantum gravity~\cite{Garay}, black holes physics~\cite{Garay,Maggiore1,Maggiore2,Maggiore3,Hossenfelder} and doubly special relativity (DSR)~\cite{Gamboa} predict the existence of a minimal measurable length and a maximal observable momentum \cite{Ali,Das}. These theories argue that near the Planck scale, the Heisenberg Uncertainty Principle should be replaced by the so called Generalized Uncertainty Principle (GUP). The commutators which are consistent with these theories are given by~\cite{Ali, Das},

\begin{equation}
\label{1}
\left [x_{i},p_{j}\right]=i\hbar\left[\delta_{ij}-\alpha\left(p\delta_{ij}+\frac{p_i p_j}{p}\right)+\alpha^2\left(p^2 \delta_{ij}+3p_i p_j\right)\right],
\end{equation}

where $p^2=\Sigma_{j=1}^{3}p_j p_j$, $\alpha=\alpha_{0}/M_{Pl}c=\alpha_{0}\ell_{Pl}/\hbar$, $M_{Pl}$ is the Planck mass, $\ell_{Pl}\approx10^{-35}$ m is the Planck length and $M_{Pl}c^2$ is the Planck energy $\approx 10^{19}$GeV.
Then, the commutation relation Eq.~\eqref{1} is approximately satisfied by following representation~\cite{Ali,Das}
\begin{equation}
\label{3}
p_{i}=p_{0i}\left(1-\alpha{p_{0}}+2\alpha^2 {p_{0}}^2\right),~~~~~ x_i=x_{0i},
\end{equation}
where $x_{i0}$ and $p_{i0}$ obey the canonical commutation relation $[x_{i0}, p_{j0}] =i\hbar\delta_{ij}$.
As it is known, the Liouville theorem says that during the time evolution, the number of quantum states inside phase space should be fixed in the presence of GUP framework. Hence, GUP should modify the density states which leads to a modification in the statistical and thermodynamical properties of any physical system. This implies, the following modification in the number of quantum states per momentum space volume as~\cite{Ali3},
\begin{equation}\label{3i}
\frac{V}{(2\pi)^3} \int_0^{\infty}d^3p \rightarrow \frac{V}{(2\pi)^3} \int_0^{\infty}\frac{d^3p}{(1-\alpha p)^4}.
\end{equation}
To obtain the thermodynamics of photons system, we need to derive the partition function. So, using Eq.~\eqref{3i}, the modified partition function per unite volume is expressed as
\begin{eqnarray}\label{3ii}
\nonumber \ln{Z} =-\frac{g_{\pi}}{2\pi^2}\int_0^{\infty}\ln\left[1-e^{-\frac{p}{T}}\right]\frac{p^2dp}{(1-\alpha p)^4},~~~~ \\
\simeq-\frac{g_{\pi}}{2\pi^2}\int_0^{\infty}\ln\left[1-e^{-\frac{p}{T}}\right](1+4\alpha p)~p^2dp,
\end{eqnarray}
where $g_{\pi}$ refers to the number of degrees of freedom. Now, the solution of (\ref{3ii}) is given by
\begin{eqnarray}\label{3iii}
\nonumber \ln{Z} =\frac{g_{\pi}}{2\pi^2}\frac{1}{T}\int_0^{\infty}\frac{1}{e^{\frac{p}{T}}-1}~\left[\frac{1}{3}p^3+\alpha p^4\right]~dp, \\
=\left[\frac{\pi^2 g_{\pi}}{90}T^3+\alpha \frac{12\zeta_5g_{\pi}}{\pi^2}T^4\right],~~~~~~~~~~~~~~~~~
\end{eqnarray}
which $\zeta_5$ denotes to the Hurwitz zeta function.
Now, with the modified partition function in hand, we can obtain the entropy of photon gas as
\begin{equation}\label{3iiii}
S=\frac{\partial}{\partial T}(T\ln{Z})=\frac{2\pi^2 g_{\pi}}{45}T^3+\alpha \frac{60\zeta_5 g_{\pi}}{\pi^2}T^4,
\end{equation}
from which we can go back to the standard entropy i.e., $S = (2g_{\pi}\pi^2/45)T^3$  by setting $\alpha\rightarrow0$.

On the other hand, It is found that the most important results of the considered GUP model  is the space discrete, or all measurable lengths are quantized in units of a fundamental minimum measurable length $\alpha=\alpha _0 l_{Pl}$. This fundamental length cannot exceed the electroweak length scale $10^{17}~l_{Pl}\approx 10^{-2}~GeV$.
The upper bound in GUP parameter has been studied in \cite{1}. In this study many quantum phenomena such harmonic oscillator, Landau levels spacing, tunnelling effect and Lamb shift are considered under the effect in GUP. They have found that the upper bounds on $\alpha_0$ lies within $10^{10}\sim 10^{23}$, which means that the GUP parameter $\alpha$  lies within $10^{4}\sim 10^{-9}$ GeV$^{-1}$.
Another phenomenological study of GUP in a gravitational phenomena are considered in \cite{2}, such as Deflection of light, time delay of light, perihelion precession, and gravitational redshift. It is found that the upper bounds on $\alpha_0$ lies within $10^{35}\sim 10^{41}$, which means that the GUP parameter $\alpha$  lies within $10^{16}\sim 10^{22}$ GeV$^{-1}$. The previous results are very greater that those reported with quantum mechanical predictions, see \cite{Ali,Das,5}.
Therefore, we find that it is not necessary to adhere to specific values of the GUP parameter as there is a very large range for its predicted value. We will just pick random values within range to show an effect of GUP in SGW signal.

%%%%%%%%%%%%%%%%%%%%%%%%%%%%%%%%%%%%%%%%%%%%%%%%%
\section{The effects of GUP modification on SGW spectrum}
\label{sec:effGUPSGW}
%%%%%%%%%%%%%%%%%%%%%%%%%%%%%%%%%%%%%%%%%%%%%%%%%
In this part, we focus on the SGW spectrum, which is agreed to be generated during the epoch of cosmological QCD phase transitions to the current period, within the framework of GUP (Eq.~\eqref{1}). With a good approximation, to examine the observable spectrum of SGW, we consider the expansion of the universe to be adiabatic ($\dot{S}/S = 0$) which leads the total entropy remains constant even beyond equilibrium.
Given that the number of photons is much higher than the number of baryons in the universe, the entropy of the universe is dominated by the photon bath. So, applying the modified entropy in (\ref{3iiii}), the relevant entropy density may be written as
\begin{equation}
S\sim a^3 g_s\left[\frac{2\pi^2}{45}T^3+\alpha \frac{60\zeta_5}{\pi^2}T^4\right],
\end{equation}
where $a$ is the scale factor and $g_s$ is the effective number of degrees of freedom involved in entropy density. Accordingly, using adiabaticity condition $\dot{S}/S = 0$, the following ansatz is given for time variation of universe temperature as
\begin{equation}\label{4i}
\frac{dT}{dt}=-\frac{H}{W(\alpha,T)},
\end{equation}
where $H$ is the Hubble parameter and
\begin{equation}
W(\alpha,T)=\frac{1}{T}\left[1+\frac{T}{3g_s}\frac{dg_s}{dT}+\frac{450 \alpha \zeta_5 T}{\pi^4+5400 \alpha \zeta_5 T}\right].
\end{equation}

Now, Eq. (\ref{4i}) in terms of scale factor is given by
\begin{equation}\label{10}
\frac{a_*}{a_0}=\exp{\left[\int_{T_*}^{T_0}W(\alpha,T) dT\right]},
\end{equation}
where the subscripts ``*" and ``0" represent the respective quantities at the epochs of phase transition and today, respectively. Then, from the relation between the scale factor and redshift, i.e. $\nu_{0\text{peak}}/\nu_{*} = a_{*}/a_0$, the redshift in the SGW frequency peak relative to the corresponding value at current epoch is expressed as
\begin{eqnarray}
\frac{\nu_{0\text{peak}}}{\nu_*}
=\frac{T_0}{T_*}\left[\frac{g_s(T_0)}{g_s(T_*)}\right]^{\frac{1}{3}}
\left[\frac{\pi^4+5400 \alpha \zeta_5 T_0}{\pi^4+5400 \alpha \zeta_5 T_*}\right]^{\frac{1}{12}}.
\end{eqnarray}
\begin{figure}
\includegraphics[width=7cm, height=5cm]{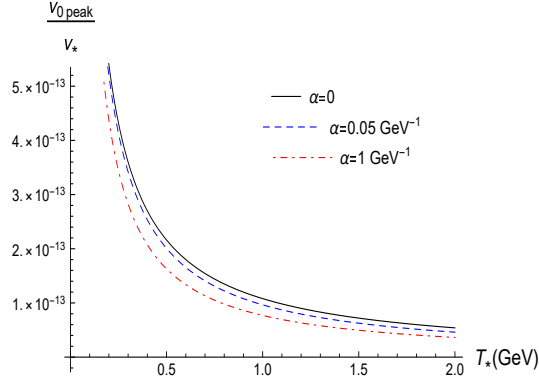}
\caption{$\nu_{0\text{peak}}/\nu_{*}$ as a function of transition temperature $T_*$ for some values of $\alpha$. In this plot, we set $T_0=2.725K=2.348\times 10^{-13}$ GeV, $g_s(T_*)\in\left[33-37\right]\approx 35$ and $g_s(T_0)=3.4$.}
\label{fig:1}
\end{figure}
To proceed further, in Fig.~\ref{fig:1} we have depicted the ratio of the frequency at the current time to that presents at the epoch of transition as a function of the transition temperature. According to this figure, for fixed values of $T_*$, $\nu_{0\text{peak}}/\nu_{*}$ decreases with increasing $\alpha$. It means that the effects of minimal length and maximal momentum reduce $\nu_{0\text{peak}}/\nu_{*}$ in comparison to its counterpart in the absence of the GUP effects. Moreover, it is shown that when the transition temperature increases, $\nu_{0\text{peak}}/\nu_{*}$ is decreased for a fixed value of $\alpha$.

We now use the Boltzmann equation, that is $\frac{d}{dt}\left(\rho_{gw}a^4\right)=0$, to obtain the energy density. This idea comes from the fact that SGW will eventually be decoupled from dynamics of the rest of the universe. So, the energy density of SGW at current time implies
\begin{eqnarray}\label{b4}
\nonumber \rho_{gw}(T_0)=\rho_{gw}(T_*) \left(\frac{a_*}{a_0}\right)^4, ~~~~~~~~~~~~~~~~~~~~~~\\
=\rho_{gw}(T_*)~\exp{\left[\int_{T_*}^{T_0}4W(\alpha,T)dT\right]}.
\end{eqnarray}
In this step, we define the density parameter of SGW at phase transition epoch as $\Omega_{gw*}=\frac{\rho_{gw}(T_*)}{\rho_{cr}(T_*)}$ and also its counterpart at current time in the shape of $\Omega_{gw}=\frac{\rho_{gw}(T_0)}{\rho_{cr}(T_0)}$ where $\rho_{cr}$ is the critical density.
From Eq. (\ref{b4}), we then obtain
\begin{equation}\label{b5}
\Omega_{gw}=\Omega_{gw*}\left(\frac{H_*}{H_0}\right)^2
\exp{\left[\int_{T_*}^{T_0}4W(\alpha,T)dT\right]},
\end{equation}
where
\begin{equation}\label{b6}
\left(\frac{H_*}{H_0}\right)^2=\frac{\rho_{cr}(T_*)}{\rho_{cr}(T_0)}.
\end{equation}
To examine the ratio of the Hubble parameter at the epoch of transition to that of its current value, we need to apply the continuity equation as
$\dot{\rho}_t=-3H\rho_t\left(1+\frac{P_t}{\rho_t}\right)$ where $\rho_t(P_t)$ refers to the total energy density (pressure) of the universe and dot implies the derivative with respect to cosmic time.
So, using Eq.~\eqref{4i}, the continuity equation in terms of temperature can be cast as

\begin{equation}\label{b8}
\frac{d\rho_t}{\rho_t}=3(1+\omega)W(\alpha,T)dT,
\end{equation}
where $\omega=\frac{P_t}{\rho_t}$ is the effective equation of state parameter with the possibility of temperature dependence.

Now, by integrating Eq.~\eqref{b8} between two intervals, the early time $T(r)$ where radiation is predominant, to the time of transition $T_*$, the critical energy density of radiation in the phase transition period $\rho_{cr}(T_*)$ can be obtained as
\begin{equation}\label{b9}
\rho_{cr}(T_*)=\rho_r(T_r)\exp{\left[\int_{T_r}^{T_*}3(1+\omega)W(\alpha,T)dT\right]}.
\end{equation}
Then, substituting $\rho_{cr}(T_*)$ from Eq.~\eqref{b9} to Eq.\eqref{b6} leads to
\begin{equation}\label{b10}
\left(\frac{H_*}{H_0}\right)^2=\Omega_{r0}\frac{\rho_r(T_r)}{\rho_r(T_0)}
\exp{\left[\int_{T_r}^{T_*}3(1+\omega)W(\alpha,T)dT\right]},
\end{equation}
where $\Omega_{r0}=\frac{\rho_r(T_0)}{\rho_{cr}(T_0)}\simeq 8.5\times 10^{-5}$ and it can be defined as the current value of fractional energy density of radiation. On the other hand, using Boltzmann equation, we can prove $\frac{\rho_r(T_r)}{\rho_r(T_0)}=(\frac{a_0}{a_r})^4$ which leads Eq.~\eqref{b10} to the following expression,
\begin{equation}\label{b13}
\left(\frac{H_*}{H_0}\right)^2=\Omega_{r0}
\exp{\left[\int_{T_0}^{T_r}4W(\alpha,T) dT\right]}
\exp{\left[\int_{T_r}^{T_*}3(1+\omega)W(\alpha,T)dT\right]}.
\end{equation}
Finally, using Eqs.~\eqref{b13} and \eqref{b5}, we obtain
\begin{eqnarray}\label{b12}
\Omega_{gw}=\Omega_{r0} \Omega_{gw*}
\exp{\left[\int_{T_*}^{T_r}4W(\alpha,T)dT\right]}
\exp{\left[\int_{T_r}^{T_*}3(1+\omega)W(\alpha,T)dT\right]}.
\end{eqnarray}
In the following subsections, we investigate the functional form of the effective equation of state for two cases of SGW with ideal gas and with QCD equation of states, respectively.
%%%%%%%%%%%%%%%%%%%%%%%%%%%%%%%%%%%%%%%%%%%%%%%%%
\subsection{SGW with ideal gas equation of state}
%%%%%%%%%%%%%%%%%%%%%%%%%%%%%%%%%%%%%%%%%%%%%%%%%
In this subsection, let us consider the ultra-relativistic gas with non-interacting particles. For this case, the effective equation of state $\omega_{\rm eff}$ is equal to $\frac{1}{3}$. So, Eqs.~\eqref{b13} and \eqref{b12} lead to,
\begin{equation}
\left(\frac{H_*}{H_0}\right)^2=\Omega_{r0}\left(\frac{T_*}{T_0}\right)^4
\left[\frac{g_s(T_*)}{g_s(T_0)}\right]^{\frac{4}{3}}\left[\frac{\pi^4+5400 \alpha \zeta_5 T_*}{\pi^4+5400 \alpha \zeta_5 T_0}\right],
\end{equation}
\begin{equation}
\Omega_{gw}=\Omega_{r0}{\Omega_{gw*}},
\end{equation}
respectively. It is clear that the transition temperature $T_*$ is always greater than the latter temperature $T_0$, so we expect that the ratio between the value of the Hubble parameter in time of phase transition and its current value increases with the comparison with its counterpart in the absence of GUP effect.
On the other hand, it is confirmed that at  temperature around a few hundred MeV the equation of state is deviated due to the role of QCD interactions ~\cite{SAN}. So QCD effect should be taken into account.
%

%

%%%%%%%%%%%%%%%%%%%%%%%%%%%%%%%%%%%%%%%%%%%%%%%
\subsection{SGW with QCD equation of state}
\begin{figure}[!htbp]
\includegraphics[width=7cm, height=5cm]{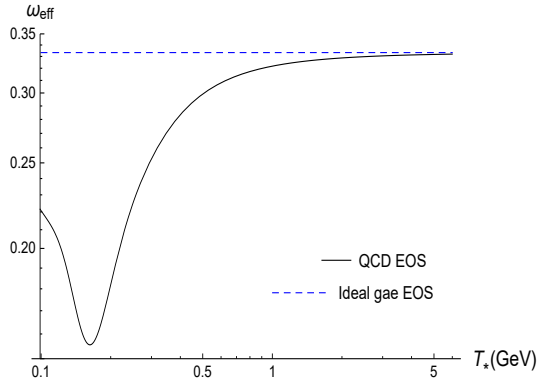}
\caption{ The effective equation of state parameter versus transition temperature $T_{*}$.}
\label{fig:eos}
\end{figure}

%%%%%%%%%%%%%%%%%%%%%%%%%%%%%%%%%%%%%%%%%%%%%%%
In this part, we want to consider the effects of QCD equation of state on SGW. The impact of QCD interaction can be introduced by employing the results of modern lattice calculation using $N_f=2+1$ flavours (it means that similar masses of $u$ and $d$ quarks and the larger mass of $s$ quark are considered) that covering a temperature range from $0.1$ GeV to $0.4$ GeV~\cite{x1}. Then, the resulting parametrization of the pressure of $u,~d,~s$ quarks and gluons in that range of temperature, is given by
\begin{equation}\label{19}
\frac{P}{T^4}=F(T)=\frac{1}{2}\left(1+\tanh{[c_{\tau}(\tau-\tau_0)]}\right)
\frac{p_i+\frac{a_n}{\tau}+\frac{b_n}{\tau^2}+\frac{c_n}{\tau^4}}
{1+\frac{a_d}{\tau}+\frac{b_d}{\tau^2}+\frac{c_d}{\tau^4}},
\end{equation}
where $\tau=\frac{T}{T_c}$ and $T_c=0.145$ GeV is the phase transition temperature and $p_i=\frac{19\pi^2}{36}$ refers to the ideal gas value of $\frac{P}{T^4}$ for QCD with three massless quarks.
\begin{table}[h!]
\centering
\begin{tabular}{ c c c c }
\hline
\hline
$c_{\tau}$&~~~~$a_n$&~~~~$b_n$&~~~~$d_n$\\
$3.8706$&~~~~$-8.7704$&~~~~$3.9200$&~~~~$0.3419$\\
\hline
$\tau_0$&~~~~$a_d$&~~~~$b_d$&~~~~$d_d$\\
$0.9761$&~~~~$-1.2600$&~~~~$0.8425$&~~~~$-0.0475$\\
\hline
\hline
\end{tabular}
\caption{Values of numerical coefficients used in Eq.~\eqref{19} to describe QCD pressure of (2+1) flavor and gluons \cite{Cheng}.}
\label{tab:1}
\end{table}
The numerical values of coefficients in Eq.~\eqref{19} are given in Table~\ref{tab:1}, for all temperatures above $100$ MeV. Also, the energy density and pressure relation can be obtained from the trace anomaly relation, namely \cite{Cheng}
\begin{equation}\label{20}
\frac{\rho-3P}{T^4}=T\frac{d}{dT}\left(\frac{P}{T^4}\right).
\end{equation}
Hence, the effective equation of state in the presence of QCD effect is given by
\begin{equation}\label{21}
\omega=\frac{P}{\rho}=\left[\frac{T}{F(T)}\frac{dF(T)}{dT}+3\right]^{-1}.
\end{equation}
In Fig.~\ref{fig:eos}, we plot the behaviour of the effective equation of state function in terms of transition temperature, $T_*$. According to this figure, around 5 GeV, the trace anomaly matches with ideal gas. It is also shown that the effect of trace anomaly near the QCD transition epoch should be considered.
%\begin{figure}
%\includegraphics[width=7cm, height=5cm]{fig2.eps}
%\caption{The effective equation of state parameter versus transition temperature $T_{*}$.}
%\label{fig:eos}
%\end{figure}
%
%***********************************************

%***********************************************
\begin{figure}[!htbp]
\includegraphics[width=7cm, height=5cm]{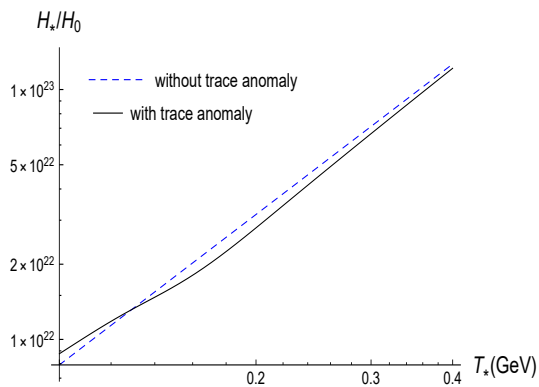}
\caption{$\frac{H_*}{H_0}$ versus $T_*$ without GUP effects for $T_r=10^4$ GeV. We set $T_r=10^4$ GeV and $g_s (T_r)=106$.}
\label{fig3}
\end{figure}

\begin{figure}[!htbp]
\includegraphics[width=7cm, height=5cm]{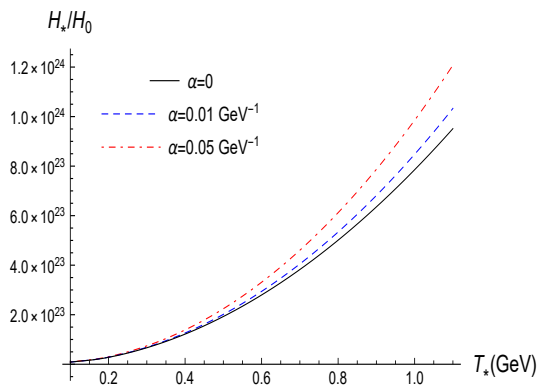}
\caption{The behaviour of $\left(\frac{H_*}{H_0}\right)$ versus transition temperature $T_*$ with $\omega_{\rm eff}$ in $QCD$ framework for some values of GUP parameter $\alpha$. We set $T_r=10^4$ GeV and $g_s (T_r)=106$.}
\label{fig:hstbyh0GUP}
\end{figure}

Now, applying  Eq. (\ref{b13}) and (\ref{21}), we can obtain
\begin{eqnarray}
\nonumber \left(\frac{H_*}{H_0}\right)^2=\Omega_{r0}\left(\frac{T_r}{T_0}\right)^4
\left[\frac{g_s(T_r)}{g_s(T_0)}\right]^{\frac{4}{3}}
\left[\frac{\pi^4+5400 \alpha \zeta_5 T_r}{\pi^4+5400 \alpha \zeta_5 T_0}\right]^{\frac{1}{3}}\times\\
\frac{g_s(T_*)^{1+\omega(T_*)}}{g_s(T_r)^{1+\omega(T_r)}}
\exp{\left[\int_{T_r}^{T_*}\frac{3(1+\omega)}{T}(1+\frac{450 \alpha \zeta_5 T}{\pi^4+5400 \alpha \zeta_5 T})dT\right]}.
\end{eqnarray}
In Fig.~\ref{fig3}, the relative Hubble parameter versus transition temperature $T_*$ with (solid line) and without trace anomaly (dashed line) is plotted for $\alpha = 0$. According to this plot, in the framework of QCD, at a transition temperature less than $0.13$ GeV, the Hubble parameter changes slowly and then, as the temperature increases, the changes become faster until it reaches to $T_*^{2}\sim 5$ GeV$^2$.
%
%\begin{figure}
%\includegraphics[width=7cm, height=5cm]{fig4.eps}
%\caption{The behaviour of $\left(\frac{H_*}{H_0}\right)$ versus transition temperature $T_*$ with $\omega_{eff}= QCD$ for some values of GUP parameter $\alpha$. We set, $T_r=10^4$ GeV and $g_s (T_r)=106$}
%\end{figure}
In Fig.~\ref{fig:hstbyh0GUP}, we have plotted the ratio between Hubble parameter at the epoch of transition to its counterpart at current epoch versus the transition temperature for different values of GUP parameter. It is shown that the effects of minimal length and maximal momentum increase the $\left(\frac{H_*}{H_0}\right)$ in comparison to its counterpart in the absence of GUP parameter, i.e., $\alpha\rightarrow0$. Also, if $\alpha_{1}> \alpha_2$ it is concluded that $\left.\frac{H_*}{H_0}\right|_{\alpha_{1}} > \left.\frac{H_*}{H_0}\right|_{\alpha_{1}}$ and these differences become bigger when temperature increases.\\
Ultimately, Eq.~\eqref{b12} can be cast in the following form,
\begin{align}
\label{25}
\frac{\Omega_{gw}}{\Omega_{gw*}} = \Omega_{r0}
&\left(\frac{T_r}{T_*}\right)^4
\left[\frac{g_s(T_r)}{g_s(T_*)}\right]^{\frac{4}{3}}
\left[\frac{\pi^4+5400 \alpha \zeta_5 T_r}{\pi^4+5400 \alpha \zeta_5 T_*}\right]^{\frac{1}{3}}\nonumber \\
&\times \frac{g_s(T_*)^{1+\omega(T_*)}}{g_s(T_r)^{1+\omega(T_r)}}
\exp{\left[\int_{T_r}^{T_*}\frac{3(1+\omega)}{T}\left(1+\frac{450 \alpha \zeta_5 T}{\pi^4+5400 \alpha \zeta_5 T}\right)dT\right]}.
\end{align}
\begin{figure}[!htbp]
\includegraphics[width=7cm, height=5cm]{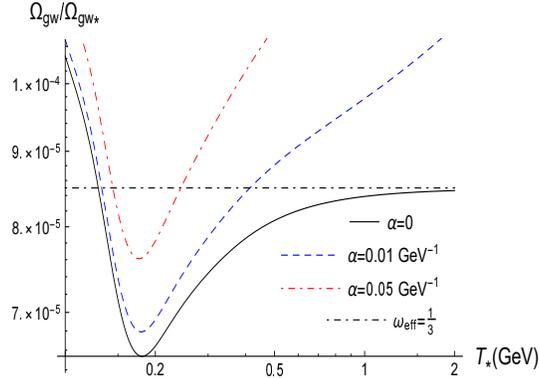}
\caption{$\frac{\Omega_{gw}}{\Omega_{gw*}}$ versus transition temperature $T_*$ with trace anomaly formalism for different values of $\alpha$. We set, $T_r=10^4$ GeV and $g_s (T_r)=106$.}
\label{fig:omgbyomgst}
\end{figure}
In Fig.~\ref{fig:omgbyomgst}, with trace anomaly equation in hand, we have depicted the relative density parameter for some values of $\alpha$ and also the equation of states of ultra-relativistic non-interacting gas is considered there. It is shown that the effects of GUP increase the density parameter ratio and it will be bigger when $\alpha$ increases. In other word, $\left.\frac{\Omega_{gw}}{\Omega_{gw*}}\right|_{\alpha=0} < \left.\frac{\Omega_{gw}}{\Omega_{gw*}}\right|_{\alpha\neq0}$ and if $\alpha_{1}> \alpha_2> \alpha_3$ then we have $\left.\frac{\Omega_{gw}}{\Omega_{gw*}}\right|_{\alpha_1} > \left.\frac{\Omega_{gw}}{\Omega_{gw*}}\right|_{\alpha_2} > \left.\frac{\Omega_{gw}}{\Omega_{gw*}}\right|_{\alpha_3}$. Also in this figure, considering equation of state for ultra-relativistic gas with non-interacting particles ($\omega_{\rm eff}=\frac{1}{3}$), leads Eq. (\ref{b12}) to a fixed line i.e., $\frac{\Omega_{gw}}{\Omega_{gw*}}=\Omega_{r0}=8.5\times 10^{-5}$ \cite{Caprini10}. Also, from the figure, when the QCD equation of state goes to the equation of state of the ultra-relativistic non-interacting gas, the relative density parameter also goes to the relative density parameter at about $2$ GeV.

%%%%%%%%%%%%%%%%%%%%%%%%%%%%%%%%%%%%%%%%%%%%%%%
\section{Modified QCD sources of stochastic gravitational wave}
\begin{figure}[t]
  \centering
  \subfloat[]{
    \includegraphics[scale=0.9]{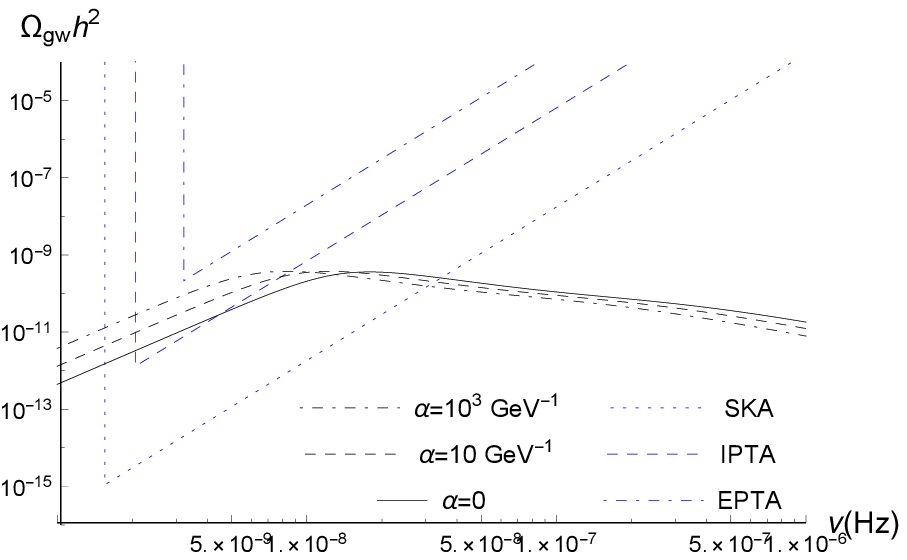}
    \label{fig:6a}
    }~~
  \subfloat[]{
    \includegraphics[scale=0.9]{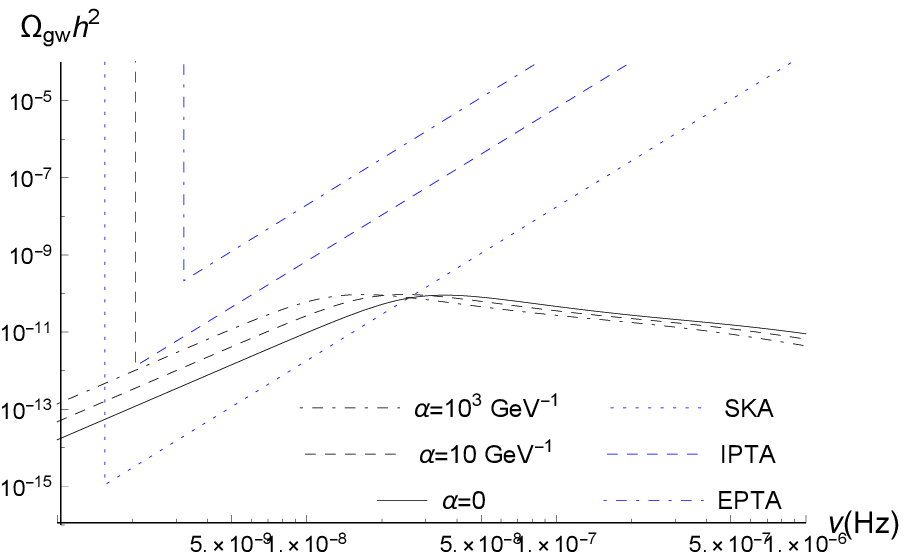}
    \label{fig:6b}
    }
\caption[]{SGW spectrum for different parameter values: (a) $n=5$, $\mu=0.7$, (b) $n=10$, $\mu=0.7$. The blue dotted, dashed, and dash-dotted lines represent the observational reaches of SKA, IPTA, and EPTA, respectively.}
\label{fig:4a4b}
\end{figure}

%%%%%%%%%%%%%%%%%%%%%%%%%%%%%%%%%%%%%%%%%%%%%%%
In this part, to determine the SGW spectrum, we focus on the density parameter of gravitational wave at the epoch of transition $\Omega_{gw*}=\Omega_{gw}(T_*)$. Although there are various sources such as solitons and solitons stars \cite{Gleiser}, cosmic strings and domain walls \cite{Vachaspati,Vachaspati2} that contribute to the SGW background, we are here interested in examining the generalized cosmological background based on the first order phase transition in the early universe. These phase transitions give rise to two significant components involved in the production of SGW at first order phase transition namely the collision of bubble walls (bwc) and shocks in the plasma \cite{CJHO,Kosowsky1,Kosowsky2,Kamionkowski,Caprini3,Huber3} magnetohydrodynamic (mhd) turbulence which may be produced after plasma's bubble collision \cite{Caprini4}.
Now, applying envelope approximation and also with numerical simulation, the contribution to the SGW spectrum by bubble collisions reads \cite{Huber3,Jinno},
\begin{equation}
\Omega_{gw*}^{bwc}(\nu)=\left(\frac{H_*}{\beta}\right)^2
\left(\frac{\kappa_b\epsilon}{1+\epsilon}\right)^2\left(\frac{0.11 \mu^3}{0.42+\mu^2}\right)\frac{3.8\left(\frac{\nu}{\nu_b}\right)^{2.8}}
{1+2.8\left(\frac{\nu}{\nu_b}\right)^{3.8}},
\end{equation}
where $\beta$ is the inverse time duration of the phase transition, $\kappa_b$ refers to the fraction of the latent heat of the phase transition deposited on the bubble wall, $\epsilon$ is the ratio of the vacuum energy density released in the phase transition to that of the radiation, $\mu$ denotes the velocity of wall and
\begin{equation}
\nu_b=\frac{0.62\beta}{1.8-0.1\mu+\mu^2}\left(\frac{a_*}{a_0}\right),
\end{equation}
is today's peak frequency of the SGW which generated by bwc mechanism during phase transition. It is believed that during the QCD phase transition, the  kinetic and magnetic Reynolds numbers of cosmic fluid are huge \cite{Caprini}. So, the percolation of the bubbles into fully ionized plasma can lead to producing of mhd turbulence. Now, using Kolmogorov-type turbulence \cite{Kosowsky}, we can express the contribution to the SGW spectrum by bubble collisions as \cite{Caprini,38},
\begin{equation}
\Omega_{gw*}^{mhd}(\nu)=\left(\frac{H_*}{\beta}\right)
\left(\frac{\kappa_m\epsilon}{1+\epsilon}\right)^{\frac{3}{2}}
\mu\frac{\left(\frac{\nu}{\nu_{mhd}}\right)^3}
{\left[1+\frac{\nu}{\nu_{mhd}}\right]^{\frac{11}{3}}\left[1+\frac{8\pi\nu}{H_*}\left(\frac{a_*}{a_0}\right)^{-1}\right]},
\end{equation}
in which $\kappa_m$ denotes the fraction of latent heat converted into the turbulence and $\nu_{mhd}$ is the current peak frequency of the SGW generated by $mhd$ at the epoch of phase transition defined as
\begin{equation}
\nu_{mhd}=\frac{7\beta}{4\mu}\left(\frac{a_*}{a_0}\right).
\end{equation}
It is known that there is still no sure way to find $\kappa$. Since, the roles of parameters $\alpha$ and $\kappa$ are very important on the definitions of peak position and the SGW signal's amplitude, we use $\beta=nH_*$, $n=5$ and $10$, $\mu=0.7$ and $\frac{\kappa_b\epsilon}{1+\epsilon}=\frac{\kappa_m\epsilon}{1+\epsilon}=0.05$ \cite{SAN,SCAP}. According to the general relation between Hubble parameter and the energy density, the Hubble parameter at transition epoch can be defined as
\begin{equation}
H_*=\sqrt{\frac{8\pi}{3m_p^2}\rho(T_*)},~~~~T_*=T_c.
\end{equation}
Taking into account Eqs. (\ref{19}) and (\ref{20}), the energy density at transition temperature can be obtained as
\begin{equation}
\rho(T_*)=T_*^5\frac{dF(T_*)}{dT_*}+3T_*^4F(T_*).
\end{equation}
Therefore, following the above definitions, it is obtained
\begin{equation}
\label{33}
\Omega_{gw*}^{bwc}(\nu)=
\frac{1.6\times 10^{-5}\left(\frac{\nu}{\nu_b}\right)^{2.8}}
{1+2.8\left(\frac{\nu}{\nu_b}\right)^{3.8}},~~~~
\nu_b=1.4\left(\frac{a_*}{a_0}\right)\sqrt{\frac{8\pi}{3m_p^2}\rho(T_*)},
\end{equation}
\begin{equation}
\label{34}
\Omega_{gw*}^{mhd}(\nu)=\frac{1.6\times 10^{-3}\left(\frac{\nu}{\nu_{mhd}}\right)^3}
{\left[1+\frac{\nu}{\nu_{mhd}}\right]^{\frac{11}{3}}
\left[1+8\pi\nu\left[\left(\frac{a_*}{a_0}\right)\sqrt{\frac{8\pi}{3m_p^2}\rho(T_*)}\right]^{-1}\right]},~~~~
\nu_{mhd}=12.5\left(\frac{a_*}{a_0}\right)\sqrt{\frac{8\pi}{3m_p^2}\rho(T_*)}.
\end{equation}
Now, we set Eqs. (\ref{33}) and (\ref{34}) into $\Omega_{gw*}h^2=\left[\Omega^{bwc}_{gw*}(\nu)+\Omega^{mhd}_{gw*}(\nu)\right]h^2$ and then use it into Eq. (26).
Next, with the above benchmark numerical parameter values, we investigate the net contribution of the SGW spectrum $\Omega_{gw} h^2$ due to bubble wall collision and MHD turbulence for some values of GUP parameter $\alpha$ in Figs.~\ref{fig:6a} and \ref{fig:6b}.
From these figures, in low frequency range, bigger than $1.5 \times 10^{-5}~Hz$, the SGW signal becomes weaker with GUP effect than in the absence of GUP. But for low frequency, smaller than $1.5 \times 10^{-5}~Hz$, the results are reversed.
In other words, if $\alpha_1>\alpha_2>\alpha_3$, for low frequency $\Omega_{gw} h^2|_{\alpha_1}>\Omega_{gw} h^2|_{\alpha_2}>\Omega_{gw} h^2|_{\alpha_3}$, but for high frequency it leads to $\Omega_{gw} h^2|_{\alpha_1}<\Omega_{gw} h^2|_{\alpha_2}<\Omega_{gw} h^2|_{\alpha_3}$.

%\begin{figure}[!tbp]
%  \centering
%  \begin{minipage}[b]{0.4\textwidth}
%    \includegraphics[width=7cm, height=5cm]{6a.eps}
%    \caption{$n=5$, $\mu=0.7$}
%  \end{minipage}
%  \hfill
%  \begin{minipage}[b]{0.5\textwidth}
%    \includegraphics[width=7cm, height=5cm]{6b.eps}
%    \caption{$n=10$, $\mu=0.7$}
%  \end{minipage}
%\end{figure}

%***********************************************
%***********************************************
 In general the GWs produced from cosmological phase transition (PT) are of quite small frequency. In such a scenario the technique of detecting GWs using pulsar timing arrays (PTA) come to rescue. Usually PTAs can reach the sensitivity in the ballpark of $10^{-9}-10^{-7}$ Hz which is just the right range relevant for the GWs produced by cosmological PTs. There a number of upcoming facilities e.g., International Pulsar Timing Array (IPTA)~\cite{IPTA:2013lea}, European Pulsar Timing Array (EPTA)~\cite{Kramer:2013kea}, Square Kilometer Array (SKA)~\cite{5136190} etc. which will play a crucial role in the study of the SGWs.
From Fig.~\ref{fig:4a4b} it is also evident that for different parameter choices the contribution on SGWB be such that it can reach the sensitivity of the upcoming GW detection facilities like SKA and IPTA. The lower frequency range of this SGW as compared to other violent sources like black-hole or neutron star mergers, makes the detection of SGWB anyway a challenging task. However, from the figure it can also be seen that in the low frequency regime ($< 10^{-8}$ Hz) the increase in the GUP parameter $\alpha$ can enhance the GW energy density and thus it can slightly augment the otherwise weak signal strength.

%\begin{figure}
%\includegraphics[width=7cm, height=5cm]{fig6.eps}
%\caption{Net contribution to the SGW due to bubble wall collision and magnetohydrodynamic turbulence %for some values of GUP parameter $\alpha$}
%\end{figure}

%
\begin{figure}[!htbp]
\includegraphics[width=7cm, height=5cm]{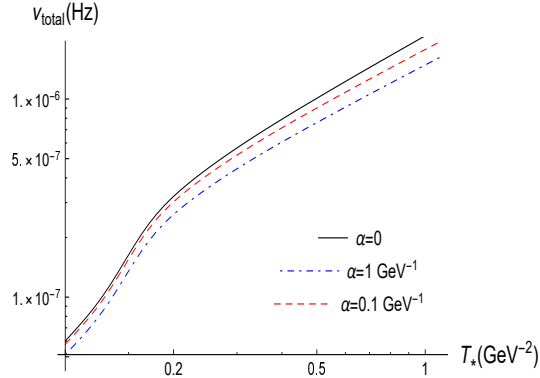}
\caption{Net peak frequency today of SGW signal arising from BWC and MHD at the epoch of phase transition for some values of $\alpha$.}
\label{fig:fig7}
\end{figure}

Finally, with the definition of total peak frequency, i.e. $\nu_{\rm total}= \nu_{bwc}+\nu_{mhd}$, from Eqs.~\eqref{33} and \eqref{34} we substitute $\nu_{bwc}$ and $\nu_{mhd}$ there which leads to
\begin{align}
\nu_{\rm total}=\left(\frac{310}{180-10v+100v^2}+\frac{35}{4v}\right)
\sqrt{\frac{8\pi}{3m_p^2}\rho(T_*)}\times
\frac{T_0}{T_*}\left[\frac{g_s(T_0)}{g_s(T_*)}\right]^{\frac{1}{3}}
\left[\frac{\pi^4+5400 \alpha \zeta_5 T_0}{\pi^4+5400 \alpha \zeta_5 T_*}\right]^{\frac{1}{12}}.
\end{align}
In Fig.~\ref{fig:fig7}, in order to make the results more clear, we have plotted the peak signal of SGW today which can be measured as a function of the transition temperature, within the GUP framework. It is indicated that, for all probable transition temperatures range, the effects of the GUP decreases the total peak frequency of the SGW signal and this difference increases with increasing transition temperature, i.e., $\nu_{\rm total}^{\alpha=0}>\nu_{\rm total}^{\alpha\neq 0}$. Also, if the effects of GUP increases, i.e., $\alpha_1 > \alpha_2 > \alpha_3$ then it leads to $\nu_{\rm total}(\alpha_1) < \nu_{\rm total}(\alpha_2)<\nu_{\rm total}(\alpha_3)$.

%%%%%%%%%%%%%%%%%%%%%%%%%%%%%%%%%%%%%%%%%%%%%%%

%%%%%%%%%%%%%%%%%%%%%%%%%%%%%%%%%%%%%%%%%%%%%%%
\section{Conclusion}
\label{sec:concl}
%%%%%%%%%%%%%%%%%%%%%%%%%%%%%%%%%%%%%%%%%%%%%%%
It is believed that during the universe evolution, it underwent various types of phase transitions. These phase transitions have different physical consequences which may be observed in our current epoch. According to some theoretical scenarios at the QCD energy scale, a first-order cosmological phase transition occurred about $t = 10^{-5}$s after the big-bang at temperature $T = 0.2$ GeV. This phase transition has a very important role in the evolution of the universe. Therefore, the study of SGW power spectrum  associated with this first-order phase transition around QGP epoch entails QGP equation of state and the trace anomaly linked with it in  an essential way because the recent lattice calculation shows that near the QCD phase transition trace anomaly has a non-vanishing contribution and it has been shown that trace anomaly reduces the expansion rate of the universe that in turn leads to an enhancement of the gravitational-wave signal.
Using QCD equation of state, a Hubble parameter, associated with the SGW, that evolved from first order QCD phase transition epoch to the current epoch has been investigated within the GUP framework. Throughout this paper, we used a GUP model, characterized with a parameter $\alpha$, which predicts two ultraviolet (UV) cutoffs namely a minimal measurable length as well as a maximum physical momentum.  We found that the GUP effects can have some impact on the SGWB. The parameter $\alpha$ can potentially increase the GW energy density at the low frequency region. Such an enhancement may play an important role as it can reach the projected sensitivities of the future generation GW detectors like SKA and IPTA.

Finally, we would like to mention that one of the current observational discrepancies in cosmology, namely the $H_{0}$ tension -- the 4.4$\sigma$ mismatch in the measurements of the Hubble constant $H_{0}$ -- can be alleviated by the indetermination associated kinematical versus dynamical measurements~\cite{Salvatore}. This finds its origin in the Heisenberg uncertainty principle in the form of a possible uncertainty in the photon mass. Now, the paradigm of GUP can be one general arena to materialise this idea. This we leave for further study in some future work.
\\

\vspace*{0.5cm}
\textbf{Acknowledgement:}
UKD acknowledges the support from Department of Science and Technology (DST), Government of India under the grant reference no. SRG/2020/000283.

%%%%%%%%%%%%%%%%%%%%%%%%%%%%%%%%%%%%%%%%%  References
%%%%%%%%%%%%%%%%%%%%%%%%%%%%%%%%%%%%%%%%%

\end{document}